# Whole-brain diffusional variance decomposition (DIVIDE): Demonstration of technical feasibility at clinical MRI systems

Filip Szczepankiewicz[1], Jens Sjölund[2,3,4], Freddy Ståhlberg[1,5,6], Jimmy Lätt[7], Markus Nilsson[5,6]

1. Lund University, Department of Clinical Sciences Lund, Medical Radiation Physics, Lund, Sweden; 2. Elekta Instrument AB, Kungstensgatan 18, Box 7593, SE-103 93 Stockholm, Sweden; 3. Department of Biomedical Engineering, Linköping University, Linköping, Sweden; 4. Center for Medical Image Science and Visualization (CMIV), Linköping University, Sweden; 5. Lund University, Skåne University Hospital, Department of Clinical Sciences Lund, Diagnostic Radiology, Lund, Sweden; 6. Lund University, Lund University Bioimaging Center, Lund, Sweden; 7. Skåne University Hospital, Department of Imaging and Function, Lund, Sweden.

**Abstract**

*Purpose:* To assess the technical feasibility of whole-brain diffusional variance decomposition (DIVIDE) based on q-space trajectory encoding (QTE) at clinical MRI systems with varying performance. DIVIDE is used to separate diffusional heterogeneity into components that arise due to isotropic and anisotropic tissue structures.
*Methods:* We designed imaging protocols for DIVIDE using numerically optimized gradient waveforms for diffusion encoding. Imaging was performed at systems with magnetic field strengths between 1.5 and 7 T, and gradient amplitudes between 33 and 80 mT/m. Technical feasibility was assessed from signal characteristics and quality of parameter maps in a single volunteer scanned at all systems.
*Results:* The technical feasibility of QTE and DIVIDE was demonstrated at all systems. The system with the highest performance allowed whole-brain DIVIDE at 2 mm isotropic voxels. The system with the lowest performance required a spatial resolution of 2.5×2.5×4 mm$^3$ to yield a sufficient signal-to-noise ratio.
*Conclusions:* Whole-brain DIVIDE based on QTE is feasible at the investigated MRI systems. This demonstration indicates that tissue features beyond those accessible by conventional diffusion encoding may be explored on a wide range of MRI systems.

**Introduction**

Diffusion MRI (dMRI) enables non-invasive imaging of tissue microstructure. Methods such as diffusion tensor imaging (DTI) (Basser et al., 1994) yield parameters which are sensitive to alterations of the tissue microstructure in both healthy development and disease (Assaf and Pasternak, 2008). However, DTI captures only a voxel-scale average, whereas quantification of diffusion heterogeneity on the sub-voxel scale warrants the use of multiple b-values and methods capable of analyzing such data. Diffusional kurtosis imaging (DKI) (Jensen et al., 2005) is an extension to DTI that can probe sub-voxel heterogeneity, which may be of importance in, for example, tumor imaging. Accordingly, several studies have found that DKI is better at differentiating high and low grade gliomas compared to DTI; presumably due to its sensitivity to tissue heterogeneity (Raab et al., 2010, Van Cauter et al., 2012). However, DTI and DKI, along with the vast majority of other dMRI techniques, are based on so-called single diffusion encoding (SDE), i.e. a pair of pulsed gradients that encode for diffusion along a single direction per acquisition (Stejskal and Tanner, 1965). Notably, methods based solely on SDE data are inherently limited since they will inadvertently conflate two separate components of heterogeneity corresponding to either microscopic anisotropy or isotropic heterogeneity (Mitra, 1995, Szczepankiewicz et al., 2016a).

A promising solution to disambiguate the two components of heterogeneity is to probe the diffusion in more than one direction in a single shot, i.e. encoding that yields b-tensors of rank two or three (Westin et al., 2016). A prominent example is double diffusion encoding (DDE), which can produce b-tensors of rank one or two and thereby disambiguate the two tissue characteristics conflated by SDE (Cory et al., 1990, Özarslan and Basser, 2008, Shemesh et al., 2010). Beyond the paradigm of pulsed gradients used in SDE and DDE, is q-space trajectory encoding (QTE), which is not constrained to pulsed gradients and can render b-tensors of rank one, two or three, e.g. linear, planar or spherical encoding (Lasič et al., 2014, Eriksson et al., 2015, Topgaard, 2016, Westin et al., 2016). We recently demonstrated that a combination of linear and spherical b-tensor encoding (LTE and STE) can be used to disambiguate microscopic anisotropy from orientation coherence by so-called diffusional variance decomposition (DIVIDE) (Szczepankiewicz et al., 2015). DIVIDE has also been used in brain tumors to separate effects of cell eccentricity and variable cell density (Szczepankiewicz et al., 2016a). In conventional DKI (Jensen et al., 2005, Jensen and Helpern, 2010), these two fundamentally different sources of diffusional variance are not separable (Szczepankiewicz et al., 2016a).

The concept of diffusional variance employed in DIVIDE, and its relation to tissue heterogeneity, can be understood by considering that, at low b-values ($b < 1$ ms/µm$^2$ = 1000 s/mm$^2$), the diffusion-weighted signal depends primarily on the apparent diffusion coefficient (ADC) (Le Bihan, 2013, Jensen, 2014). The ADC reflects the average apparent diffusivity in a voxel, and is sensitive to a range of different microstructure features, for example, anisotropy in white matter (Basser and Pierpaoli, 1996) or the cell density in tumors (Chen et al., 2013). At moderate b-values ($b < 3$ ms/µm$^2$), the signal will depend on the average diffusivity as well as the heterogeneity of the intra-voxel distribution of diffusion coefficients (DDC) (Callaghan and Pinder, 1983), i.e. the so-called diffusional variance (Yablonskiy et al., 2003, Jensen et al., 2005). By employing multiple b-tensors, e.g. LTE and STE, it is possible to decompose the diffusional variance into components caused by isotropic and anisotropic features of the microstructure (Lasič et al., 2014, Szczepankiewicz et al., 2015, Szczepankiewicz et al., 2016a). Figure 1 shows how the diffusion-weighted signal depends on anisotropic and isotropic diffusional variance when using LTE and STE.

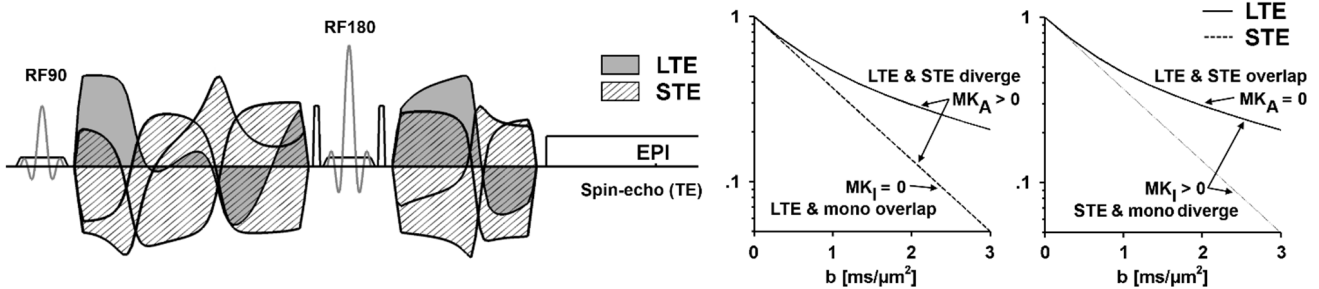

Figure 1 | Schematic spin-echo sequence with echo planar imaging (EPI) readout (left), and signal vs. b curves in two toy examples (right). The pulse sequence depicts gradients that yield linear and spherical b-tensor encoding (LTE and STE). The gradients can be asymmetric around the refocusing pulse (Sjölund et al., 2015). In the first toy example (left signal plot), we assume a tissue comprising randomly oriented and anisotropic diffusion tensors (microscopic anisotropy). In the second example, it comprises isotropic tensors with two distinct diffusivities (isotropic heterogeneity). Even if the underlying tissue is markedly different, the LTE signal is the same in both examples. However, they can be distinguished by also using STE. In the first case, the LTE and STE signal diverge (hallmark of anisotropic variance), and the STE signal is monoexponential (no isotropic variance). In the second case, the LTE and STE signals overlap (no anisotropic variance) and the STE signal is non-monoexponential (hallmark of isotropic variance).

The drawback of encoding with b-tensors of rank above one is that they are generally more demanding with respect to hardware performance, compared to conventional SDE. As such, these techniques have been developed and used primarily at high-performance MRI systems that facilitate sufficient data quality and acceptable acquisition times. For example, the initial implementation of DIVIDE was performed at a 3 T scanner with 80 mT/m gradients. Even so, it required an echo time of 160 ms to achieve a b-value of 2.8 ms/µm$^2$, which limited the spatial resolution and coverage (Szczepankiewicz et al., 2015). Due to the low signal-to-noise ratio imposed by long echo times, the technical feasibility for whole-brain DIVIDE at lower field strengths and/or gradient amplitudes, in clinically acceptable times, is not obvious. However, recent development of numerically optimized waveforms has facilitated improved encoding efficiency enabling QTE at substantially shorter echo times (Sjölund et al., 2015). This development allows for faster acquisitions and higher SNR, and motivates an investigation of the technical feasibility of QTE and DIVIDE on a wider range of systems.

The aim of this study was to survey the technical feasibility of QTE and whole-brain DIVIDE at a wide range of MRI systems, covering field strengths between 1.5 and 7 T, and gradient amplitudes between 33 and 80 mT/m. We expected a nearly linear dependency of SNR on field strength (Moser et al., 2012). However, higher SNR is not guaranteed at 7 T because the transversal relaxation time is substantially shorter which may decrease the SNR given long enough echo times (Cox and Gowland, 2010, Choi et al., 2011, Szczepankiewicz et al., 2016b). Similarly, we expected the SNR to increase with gradient performance, since stronger gradients results in shorter echo times for any a given b-value. The results showed that whole-brain DIVIDE based on QTE is feasible at all investigated systems, although the spatial resolution should be reduced at some systems to accommodate the lower SNR resulting from low field strength and/or gradient amplitude.

**Methods**

We investigated the technical feasibility of DIVIDE at MRI systems with multiple field strengths ($B_0$ = 1.5, 3, and 7 T) and maximal gradient amplitudes ($G_{max}$ = 33, 45, 60, and 80 mT/m). An overview of the systems is found in Table 1. Note that systems A and B are the same physical scanner, but that system A was artificially limited to yield a maximal gradient amplitude of 33 mT/m.

The QTE sequence was deemed technically feasible if it produced data with whole-brain coverage (120 mm contiguous coverage in feet-head direction), in-plane resolution of 2×2 mm², slice thickness of 4 mm, SNR above 3 at the highest b-value, and DIVIDE parameter maps without prominent artefacts, in less than 8 minutes.

The software used for the model fitting and waveform optimization was implemented in Matlab (The Mathworks, Natick, MA) and is available at https://github.com/markus-nilsson/md-dmri.

*Protocol design and data acquisition*

The protocols were designed using heuristic guidelines in a procedure that followed three steps: (i) determine the minimal and maximal b-values, (ii) determine the minimal echo and repetition times (TE and TR), as well as the maximal number of samples, and (iii) distribute the samples across b-tensors, b-values and encoding directions. Details for each step are given in the appendix. Protocols were adapted for each system, and are summarized in Table 1. Independent of MRI system, the following imaging parameters were kept constant: acquisition matrix 256×256×30, partial Fourier 0.75, strong fat suppression, parallel acceleration factor 2, bandwidth 1400 Hz/voxel. Notably, the default spatial resolution was 2×2×4 mm³. The impact of anisotropic voxels is commented on in the discussion.

QTE with optimized waveforms was used to achieve high encoding efficacy, and therefore high SNR. The QTE sequence was based on the spin-echo diffusion-weighting sequence provided by Philips Healthcare (Best, the Netherlands) and Siemens Healthcare (GmbH, Erlangen, Germany), and developed in the pulse programming environment provided by each vendor. Gradient waveforms were optimized based on the specifications of each MRI system (Sjölund et al., 2015). Briefly, the optimization used the Euclidean norm to create rotatable waveforms, a heat dissipation factor of η = 0.9, and a slew rate limitations of 100 T/m/s to reduce peripheral nerve stimulation (Ham et al., 1997, Hebrank and Gebhardt, 2000). The LTE waveform was calculated from the STE waveform, such that it had the same q-vector magnitude as a function of time, according to Lasič et al. (2014). An example of a spin-echo sequence with an optimized gradient waveform is shown in Figure 1.

All data was acquired in the same healthy volunteer (male, 35 y) on all systems. All experiments were approved by the Regional Ethical Review Board, and informed consent was obtained prior to participation. Diffusion weighted data was corrected for motion and eddy-currents in Elastix (Klein et al., 2010) using an extrapolation-based reference volume (Nilsson et al., 2015). No smoothing of data was used, in order to retain the impact of signal noise on the images.

All protocols yielded whole-brain coverage in under 8 minutes. However, the system with lowest performance (system A) did not pass our criteria for technical feasibility due to low SNR. Thus, we investigated the necessary reduction in spatial resolution that would compensate for the low SNR. For this protocol, the matrix size was 102×102×30, and the bandwidth was reduced to 1150 Hz/voxel to yield the same readout time. Conversely, at the system with highest performance (system D), we investigated the feasibility of increasing the resolution to 2 mm isotropic voxels, changing only the repetition time to 8400 ms, yielding an acquisition time of 16 minutes.

Table 1 – Hardware specification, imaging protocols, and quality parameters at all systems. Systems A-D used 20-channel receive head/neck-coil arrays. System E used a 32 channel transmit/receive head-coil array. All protocols used equidistant b-values in the interval [0.1, 2] ms/µm². For example, four b-shells yield $b \approx$ 0.1, 0.7, 1.4, and 2 ms/µm².

| | | A[†] | B[†] | C[‡] | D[*] | E[**] |
|---|---|---|---|---|---|---|
| *Hardware* | | | | | | |
| $B_0$ | T | 1.5 | 1.5 | 3.0 | 3.0 | 7.0 |
| $G_{max}$ | mT/m | 33 | 45 | 45 | 80 | 60 |
| $S_{max}$ | T/m/s | 125 | 200 | 200 | 200 | 100 |
| Coil-Ch | | 20 | 20 | 20 | 20 | 32 |
| *Protocol* | | | | | | |
| TE | ms | 138 | 120 | 120 | 91 | 99 |
| TR | ms | 5600 | 5000 | 5000 | 4200 | 5000 |
| $\delta_1, \delta_2$ | ms | 60, 51 | 51, 43 | 51, 43 | 35, 26 | 41, 32 |
| b-shells | | 3 | 4 | 4 | 5 | 4 |
| $n_{dir}$ | | 7, 12, 20 | 6, 6, 12, 19 | 6, 6, 12, 19 | 6, 6, 10, 11, 20 | 6, 6, 12, 20 |
| $T_{tot}$ | min | 7:50 | 7:50 | 7:50 | 7:58 | 8:00 |
| *Quality* | | | | | | |
| $Q_3$ | % | 88 | 97 | 99 | 100 | 95 |
| $Q_6$ | % | 12 | 29 | 69 | 94 | 69 |

[†] Siemens Magnetom Aera; [‡] Siemens Magnetom Skyra; [*] Siemens Magnetom Prisma; [**] Philips Achieva. $B_0$ main magnetic field strength; $G_{max}$ maximum gradient amplitude; $S_{max}$ maximum gradient slew rate; TE echo time; TR repetition time; $\delta_1$ and $\delta_2$ duration of waveform before and after refocusing pulse; $n_{dir}$ number of directions in each b-shell; $T_{tot}$ total acquisition time including preparation stage; $Q_3$ and $Q_6$ fraction of tissue where SNR is above 3 and 6 at $b$ = 2 ms/µm².

*Parameter estimation*
DIVIDE employs a joint analysis of LTE and STE data to disentangle the anisotropic and isotropic diffusional variance ($V_A$ and $V_I$). We used the inverse Laplace transform of the gamma distribution as a signal model (Jensen and Helpern, 2010, Röding et al., 2012), in accordance with previous implementations (Lasič et al., 2014, Szczepankiewicz et al., 2015). The gamma distribution is described by two parameters that correspond to the expected value (MD) and variance ($V_{DDC}$) of the DDC {Jensen, 2010 #231}, which gives an expression for the signal, according to

$$S(b, b_\Delta) = S_0 \cdot \left(1 + b \cdot \frac{V_{DDC}}{MD}\right)^{-\frac{MD^2}{V_{DDC}}}, \quad \text{Eq. 1}$$

where $S_0$ is the signal at $b = 0$ ms/μm², and the observed diffusional variance ($V_{DDC}$) depends on the b-tensor anisotropy ($b_\Delta$),

$$V_{DDC} = V_I + b_\Delta^2 \cdot V_A, \quad \text{Eq. 2}$$

with $b_\Delta = 0$ for STE and $b_\Delta = 1$ for LTE (Eriksson et al., 2015, Topgaard, 2016). To obtain rotation invariant parameters in anisotropic samples, such as neural tissue, we performed so-called 'powder averaging' (Bak and Nielsen, 1997, Edén, 2003). The powder averaged signal ($\bar{S}$) is defined as the arithmetic average of the diffusion-weighted signal across all diffusion encoding directions in a given b-shell. The encoding directions are uniformly distributed on the half-sphere using an electrostatic repulsion model (Jones et al., 1999). Parameters were estimated by fitting Eq. 1 to LTE and STE data (Lasič et al., 2014). The fitting procedure was not constrained to yield positive values for MD, $V_I$ or $V_A$. Due to the variable density of data in each b-shell, we weighted the fitting to account for the heteroscedasticity of the powder averaged signal. The diffusional variance was reported in terms of a normalized metric, defined according to

$$MK_x = 3 \cdot \frac{V_x}{MD^2}, \quad \text{Eq. 3}$$

where 'x' denotes the isotropic, anisotropic or total diffusional variance, respectively. Westin et al. (2016) referred to these components as bulk, shear, and total kurtosis. Here, we refer them as components of the diffusional variance, since they reflect the variance of the DDC (Yablonskiy and Sukstanskii, 2010). Finally, we calculated the microscopic fractional anisotropy (μFA), according to

$$\mu FA = \sqrt{\frac{3}{2}} \cdot \left(1 + \frac{MD^2 + V_I}{\frac{5}{2}V_A}\right)^{-1/2}, \quad \text{Eq. 4}$$

which can be interpreted as the FA that would be observed if all structures in the sample were aligned in parallel (Szczepankiewicz et al., 2016a, Westin et al., 2016). This means that the μFA captures the diffusional anisotropy even in samples that are isotropic on the voxel scale (Jespersen et al., 2013, Lawrenz et al., 2015, Szczepankiewicz et al., 2015).

*Analysis of repeatability*
Repeatability was tested by acquiring data twice on systems A and D. On system A, the acquisitions were performed within the same session; on system D, acquisitions were separated by 20 days. Spatial correspondence across acquisitions was achieved by transforming the raw dMRI data into pre-MNI space, where data was resampled to 1.5 mm isotropic voxels to reduce smoothing effects from the interpolation. The transform to pre-MNI space was obtained from an affine registration of an FA volume (based on LTE with $b < 1.2$ ms/μm²), to the FMRIB58 FA volume distributed in FSL (http://fsl.fmrib.ox.ac.uk/). We refer to this as the pre-MNI space, since only a linear transform was applied. To gauge the repeatability between the first and second acquisitions, we calculated the standard deviation (SD) of the difference between parameters calculated from repeated acquisitions, i.e. SD($X_1 - X_2$), where $X$ denotes the parameter values in all included voxels (Bartlett and Frost, 2008). To avoid inflated uncertainty due to partially misaligned cortical regions, only voxels in the white matter were included by excluding voxels where μFA < 0.7.

*Signal to noise estimation*
At very low SNR, the signal will be positively biased due to the rectified noise floor, with detrimental effects on parameter estimates (Jones and Basser, 2004, Dietrich et al., 2008). Since the STE was repeated several times for each b-value, we could estimate the SNR at each b-value by computing the ratio between the mean of the STE signal and its standard deviation. Assuming that the distribution of noise is approximately Rician, the noise floor bias is most prominent in regions where SNR < 3, as described by Gudbjartsson and Patz (1995). We use this threshold to detect regions where the noise floor is likely to bias the parameter estimation. As an additional gauge of data quality, we calculated the fraction of the brain volume where SNR was above 3 and 6 at $b = 2$ ms/μm², denoted $Q_3$ and $Q_6$. Voxels where MD > 1.5 µm$^2$/ms were excluded to avoid partial volume effects with cerebrospinal fluid.

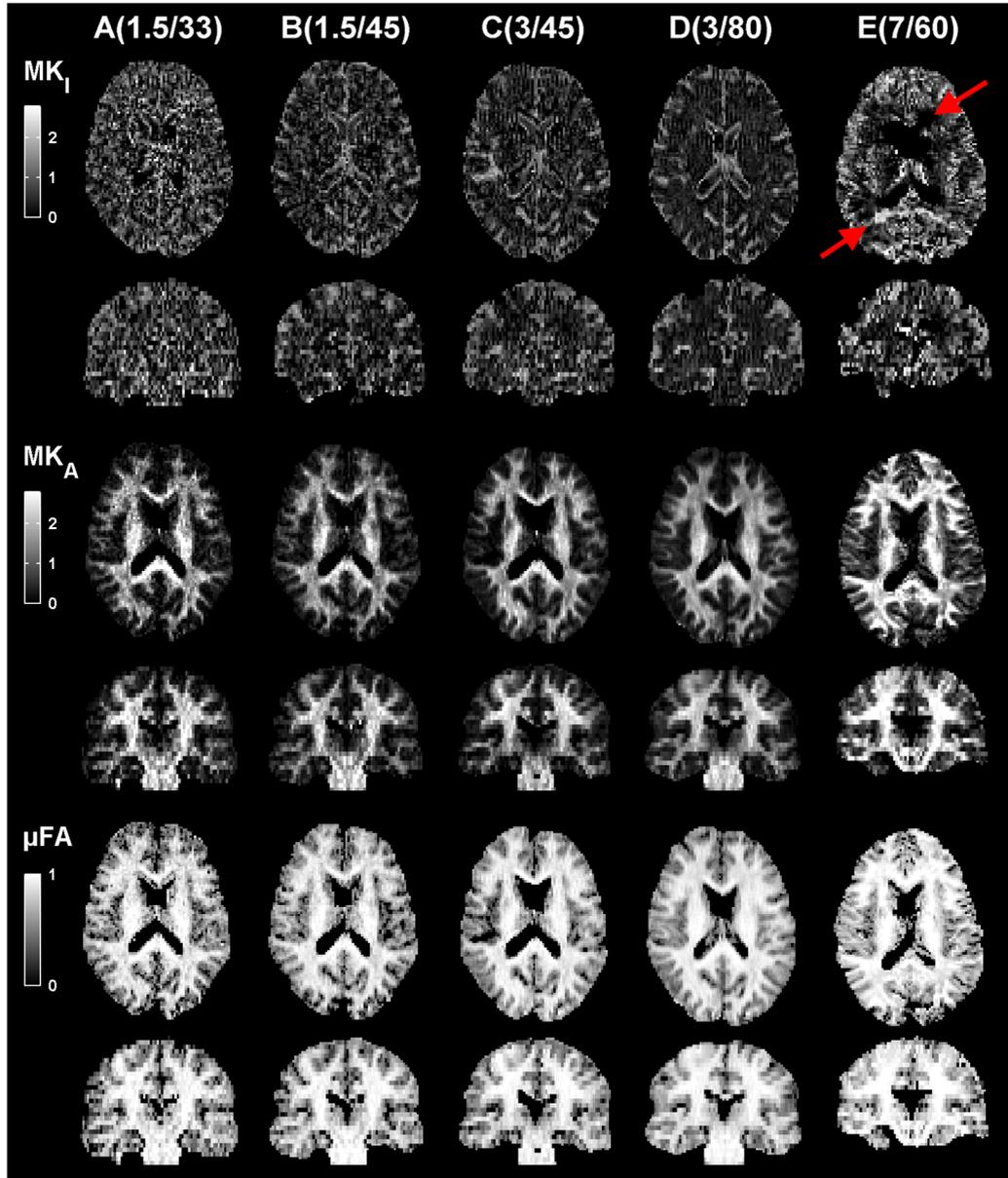

Figure 2 | Parameter maps in transversal and coronal slices, calculated from data acquired at five MRI systems (A-E, see Table 1) in one healthy volunteer. Generally, the image quality improves with field strength and gradient amplitude. Most notably there is a discernible increase in noise for systems A and B, as well as an apparent bias in the $MK_I$ map. At the 7 T system (E), two notable artifacts appeared (red arrows); the anterior region exhibited negative $MK_I$, and in the posterior region we observed a prominent fat artefact. Furthermore, system E generally exhibited higher values of $MK_A$ and µFA compared to the other systems.

*Simulation of parameter bias and precision*
We investigated if noise could account for the observed parameter accuracy. The parameter bias and precision were investigated by simulations using three different sets of tissue parameters. All three sets had MD = 1 µm$^2$/ms. The first set had $MK_I$ = 0.3 and $MK_A$ = 1.8, similar to the values detected in white matter at system D. The second set used reversed values ($MK_I$ = 1.8 and $MK_A$ = 0.3); no such values were detected in the healthy brain, but they may appear in tumor tissue (Szczepankiewicz et al., 2016a). The third set assumed that $MK_I$ = 0.0 and $MK_A$ = 1.8 to test if signal noise can account for the non-zero $MK_I$ that was detected in the white matter. The signal for each tissue was calculated from Eq. 1, and Rice distributed noise was added using the SNR estimated at each system. To represent a relatively noisy region, we estimated the SNR by placing an ROI in the corticospinal tract in an axial slice at the level of the lateral ventricles. Each system used the protocol described in Table 1 and the fitting was identical to that used in vivo, and was iterated 300 times using independent realizations of noise. The mean parameter across all iterations was calculated, along with the standard deviation and standard error of the mean.

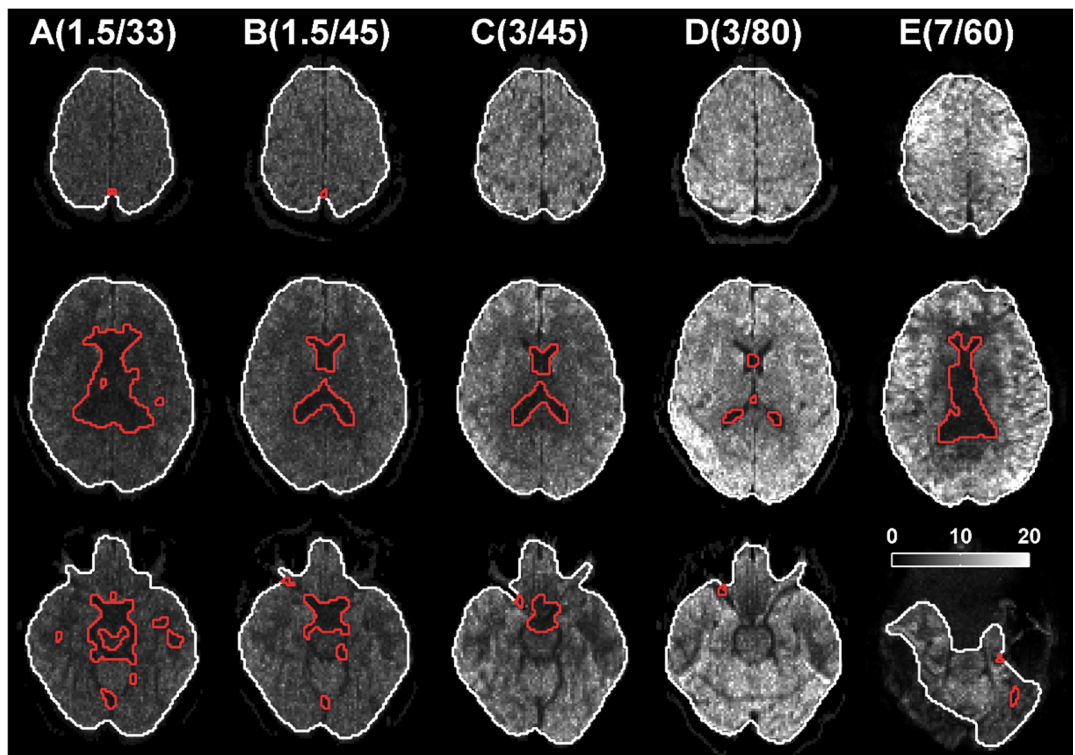

Figure 3 | SNR maps at $b = 2$ ms/µm$^2$ in three transversal slices, calculated from data acquired at five MRI systems (A-E, see Table 1) in one healthy volunteer. Red outlines show regions where SNR < 3; white outlines show the outer limit where SNR > 3. On systems B-D the SNR was only low in regions with cerebrospinal fluid. System A showed low SNR in the central and inferior parts of the brain. System E generally exhibited high SNR in the peripheral regions and low SNR in proximity to the lateral ventricles and in the lower parts of the brain. The irregular brain outline in the lower part for system E was likely caused by poor RF homogeneity (Moser et al., 2012).

## Results

QTE was performed on all five systems, according to the protocols in Table 1, and Figure 2 shows the calculated DIVIDE parameter maps. There is an appreciable difference in image quality depending on hardware performance; as expected, higher $B_0$ and higher $G_{max}$ rendered higher quality images. The quality of the parameter maps was poorest on system A, where the influence of noise was evident in all parameters. This is likely due to the extended echo time (TE = 138 ms) incurred by the relatively low $G_{max}$ of 33 mT/m. The parameter maps on system E ($B_0 = 7$ T) exhibited slightly different parameter contrast and were also noticeably affected by noise, and several image artifacts.

Figure 3 shows the estimated SNR at the highest b-value ($b = 2$ ms/µm$^2$) in three axial slices for all systems. The red outlines show regions where SNR < 3. Generally the SNR was highest in the peripheral parts of the brain, and relatively low in proximity to the lateral ventricles and in the inferior regions of the brain. Systems B-D exhibited SNR > 3 in a majority of the brain ($Q_3 > 95\%$). This was true also for system E, but only after regions with gross signal loss were excluded. Such signal dropout was likely explained by RF inhomogeneity, similar to that reported by Sigmund and Gutman (2011). As expected, system A exhibited the lowest overall SNR, and did not achieve feasible SNR levels in the central parts of the brain ($Q_3 = 88\%$ and $Q_6 = 12\%$). A complete list of the quality parameters are reported in Table 1.

Figure 4 shows the simulated parameter estimation. The SNR at $b = 0$ ms/µm$^2$ was estimated to be 16, 22, 32, 53, and 34 at systems A-E, respectively. The results show an association between parameter precision and hardware performance, and that signal noise is a plausible cause for the parameter variations seen in the parameter maps. The precision in MK$_I$ was consistently lower than for MK$_A$, regardless of their true values. Moreover, the simulations suggested that systems A and B exhibit a positive bias in MK$_I$, where the bias was 0.2 and 0.1, respectively. Remaining systems showed negligible bias (MK$_I$ and MK$_A$ differed less than 0.04 from the true value). The simulations where the true MK$_I$ was set to zero indicated that noise could not account for non-zero MK$_I$ at high SNR (system D), thus we attribute it to the presence of actual isotropic diffusional variance.

Data was also acquired on system A and D at resolutions of 2.5×2.5×4 mm$^3$ and 2×2×2 mm$^3$, to address the low SNR at system A and to test the feasibility of a higher spatial resolution at system D. Figure 5 shows the resulting parameter maps and SNR maps. For system A, the reduction in spatial resolution resulted in improved SNR, where $Q_3 = 99\%$ and $Q_6 = 61\%$ compared to 88 and 12% for the original acquisition with $2\times2\times4$ mm$^3$ voxels. For system D, the increase in spatial resolution reduced the overall SNR, but the SNR remained high in most parts of the brain ($Q_3 = 98\%$ and $Q_6 = 59\%$). These results indicate that protocols with image resolutions tailored to systems with low or high performance are also technically feasible.

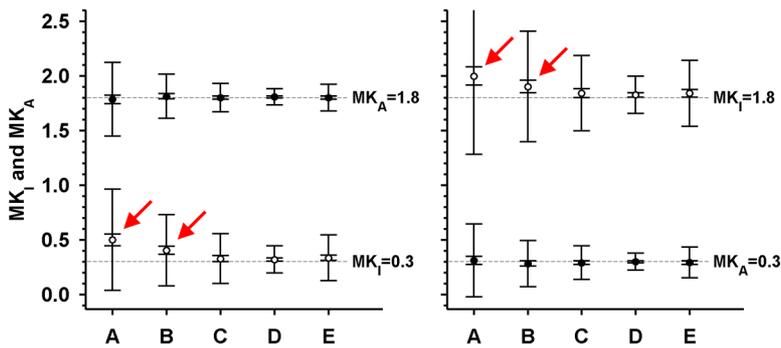

Figure 4 | Simulated effect of noise in two model tissues. Markers show the mean value; the outer whiskers show the standard deviation, and the inner whiskers show the standard error of the mean across 300 iterations of the simulation. The dashed horizontal lines show the true values used in the simulation. As expected, the parameter precision was highest for the systems that exhibited the highest SNR. Systems A and B exhibited positive bias in $MK_I$ of 0.2 and 0.1 in both sets of tissue parameters (red arrows). The bias in $MK_A$ was negligible at all systems.

Figure 6 shows the repeatability analysis for systems A and D. The parameter maps for the first and second acquisition are qualitatively similar; parameter agreement is depicted as the difference between the first and second acquisition, and in Bland-Altman plots. The repeatability was higher on system D, and the $MK_I$ consistently exhibits the highest uncertainty, in agreement with the simulations in Figure 4.

**Discussion**

We have demonstrated that DIVIDE with QTE is technically feasible across a wide range of different MRI scanners, and we have shown examples of the parameter maps that can be expected at each system for acquisition times of 8 minutes. As expected, the system with strongest gradients (system D) rendered the highest SNR and image quality. However, the technical feasibility was also demonstrated at systems B-E, where at least 95% of the brain had SNR > 3 (Table 1). By contrast, system A yielded the poorest SNR and image quality, where only 80% of the brain had SNR > 3. Accordingly, the parameter maps at system A were affected by the rectified noise floor. Additionally, the simulations indicated that system B may also exhibit a relevant parameter bias (Figure 4), especially in the central parts of the brain where SNR tended to be lower compared to peripheral tissue. Thus, the demand on SNR > 3 may not be sufficient to avoid noise-floor effects in $MK_I$. Conversely, the simulations suggested that noise contributed negligible bias in $MK_A$, for all systems. Although system A initially failed to reach our criterion for feasibility, we investigated the SNR and image quality also at a reduced spatial resolution. This yielded a feasible protocol with improved SNR and parameter map quality (Figure 5). Similarly, the superior SNR at system D facilitated a higher spatial resolution of 2 mm isotropic voxels, albeit at a prolonged scan time (Figure 5).

This study aimed at representing a wide range of MRI scanners, but hardware features other than those investigated herein may also affect image and parameter quality. For example, we cannot account for the effects of RF system, field heterogeneity, image acquisition and reconstruction, post-processing, and fitting procedure (Choi et al., 2011, Polders et al., 2011, Sigmund and Gutman, 2011, Moser et al., 2012). Furthermore, we acknowledge that the current results are based on a single healthy volunteer and may therefore lack generalizability to larger cohorts. However, we did demonstrate two aspects of technical robustness (Andersson, 1997). The test-retest data in Figure 6 demonstrated that our approach yields repeatable results within a single subject. Furthermore, we have previously demonstrated the technical robustness of the method in a proof-of-concept implementation (Sjölund et al., 2015), where QTE was performed in ten healthy volunteers at system C. Although that study used limited coverage (44 mm feet-head) and longer echo time (TE = 130 ms), the experiments yielded homogeneous data quality where $Q_3 = 98 \pm 1\%$ and $Q_6 = 52 \pm 10\%$, similar to the current quality estimated at system C (Table 1). We consider also the present implementation at 7 T to be robust (system E), although only after exempting limitation that are characteristic to ultra-high fields, such as RF inhomogeneity, and unreliable fat suppression (Sigmund and Gutman, 2011, Moser et al., 2012). Moreover, our results indicate that $MK_A$ was higher at 7 T compared to lower fields. The impact of this result on future studies is unclear and cannot be contrasted to literature since, to the best of our knowledge, no previous study has employed DKI or b-tensors of rank two or three at 7 T in vivo.

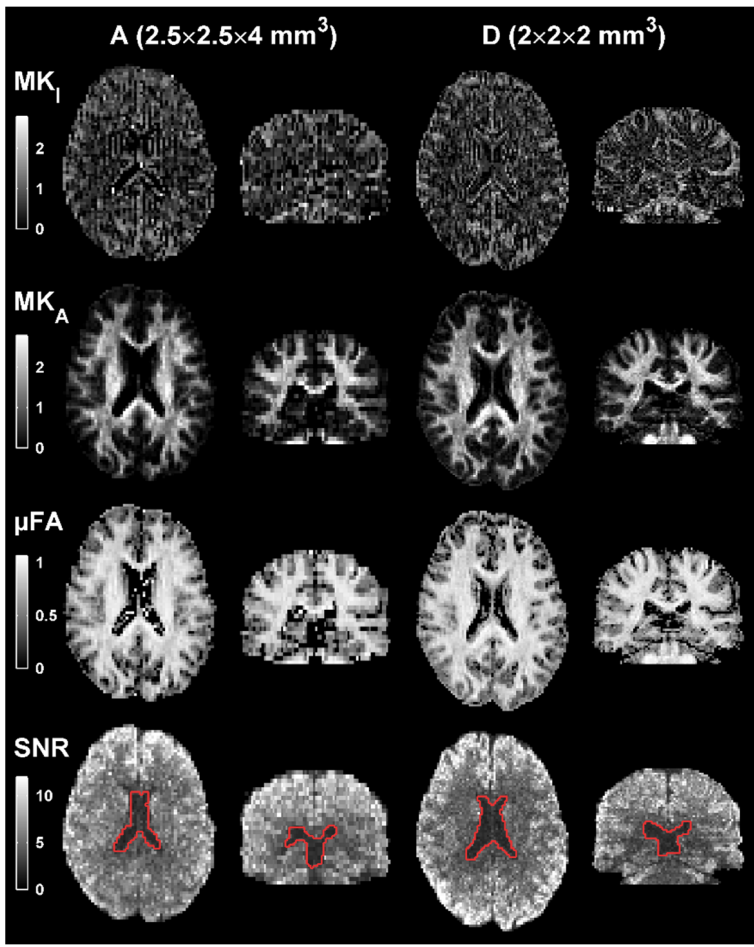

Figure 5 | Parameter and SNR maps at adjusted spatial resolution for systems A and D. The parameter maps from system A are markedly less noisy than at the initial resolution (Figure 2), and the SNR is above 3 in all regions where MD < 1.5 μm$^2$/ms ($Q_3$ = 99%, $Q_6$ = 61%). At system D the resolution was increased to 2 mm isotropic voxels, while maintaining sufficient SNR ($Q_3$ = 98%, $Q_6$ = 59%).

The feasibility of DIVIDE based on QTE hinges mainly on the use of optimized asymmetric gradient waveforms. The benefit of using the currently proposed waveforms and protocols can be appreciated by comparing them to the initial implementation which achieved a slice coverage of 15 mm in 10 minutes at an MRI scanner similar to system D (Szczepankiewicz et al., 2015). We estimate that using the previous waveform design at systems A, D, and E would render echo times of approximately 230, 140 and 170 ms, respectively. This translates to a relative SNR of 40%, 50% and 30%, assuming the white matter relaxation times presented by Cox and Gowland (2010). Asymmetric encoding has been explored previously for LTE and STE (Mori and van Zijl, 1995, Wong et al., 1995, Özarslan and Basser, 2008, Froeling et al., 2015), however, one should be aware that asymmetrical encoding also makes the imaging sequence sensitive to non-linear distortions of the gradient field, either due to non-linear gradient output or due to non-linear concomitant gradients (Bernstein et al., 1998, Baron et al., 2012). We did not observe any artifacts associated with concomitant gradient fields, however, they may impede accurate encoding at large field of views or with asymmetric gradient coils (Meier et al., 2008).

We emphasize that although this study demonstrates technical feasibility on a general level, investigation of the *clinical* feasibility, pertaining to specific applications, was outside the scope of this work. The present protocol design was intended for investigations of healthy brain tissue, and therefore accounts for the approximate diffusivity and anisotropy of brain tissue, as described in the Appendix. Naturally, a tissue with vastly different characteristics may require a different protocol. For example, a protocol designed for gliomas that exhibit MD ≈ 1.8 μm$^2$/ms and FA ≈ 0.1 (Szczepankiewicz et al., 2016a) would generally entail a lower maximal b-value ($b_{max}$ < 1.3 ms/μm$^2$) and fewer diffusion encoding directions ($n_{min}$ = 3); a markedly different premise for optimization compared to normal brain tissue (Appendix). A comprehensive protocol design should also consider the relaxation characteristics of the tissue. For example, transversal relaxation rates may depend on age and iron content, which may impact the SNR (Gelman et al., 1999, Siemonsen et al., 2008). Specific applications can also benefit substantially from adapting the image quality and scan time to the expected level of variance in the studied population (Szczepankiewicz et al., 2013), and we anticipate that simultaneous multi-slice

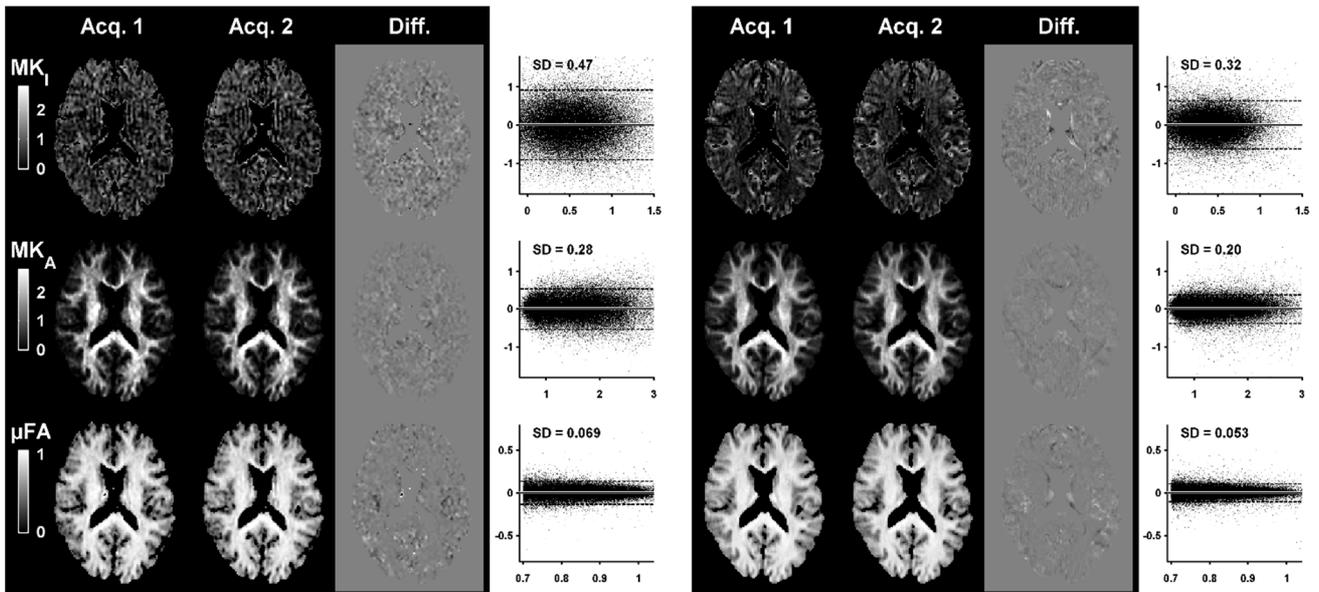

Figure 6 | Parameter maps from repeated experiments. The spatial resolution at system A was 2.5×2.5×4 mm$^3$ (left), and at system D it was 2×2×4 mm$^3$ (right). The repeatability is shown in terms of the voxel-wise difference (Diff.) between the first and second acquisition (Acq. 1 and 2). Bland-Altman plots are shown for each parameter; each data point represents a co-registered pair of voxels within the white matter. The x and y-axis of the plots show the difference and average, respectively; dashed lines show ± 2·SD($X_1 - X_2$). Note that the SD pertains to voxel-wise parameter estimation. Analysis of the average across a larger region of interest can significantly improve the parameter precision.

techniques will provide a marked improvement of acquisition times (Setsompop et al., 2012).

We wish to highlight two limitations of the present study. First, we used asymmetric voxels (2×2×4 mm$^3$) as an effective means to increase the SNR and coverage of the protocols. Asymmetric voxels are known to introduce confounding effects in measures of voxel-wise diffusion anisotropy, for example in DTI, due to the interaction between voxel and structure geometry (Oouchi et al., 2007). By contrast, DIVIDE renders parameters that are independent of the orientation coherence of the underlying tissue, thus avoiding this limitation entirely (Lasič et al., 2014). However, other issues pertaining to asymmetric voxels remain, such as increased partial volume effects in the through-plane direction. Secondly, the DIVIDE approach assumes Gaussian diffusion in each micro-environment, i.e. that the diffusion is independent of the diffusion time. The diffusion process has been shown to be time-dependent in biological tissues using specialized diffusion encoding schemes (Stanisz et al., 1997, Assaf et al., 2000, Does et al., 2003, Fieremans et al., 2016), and in pathological tissue (Lätt et al., 2009). By contrast, studies performed at more conventional diffusion times in healthy tissue indicate that the effects are negligible (Clark et al., 2001, Ronen et al., 2006, Nilsson et al., 2009). The interplay between diffusion-time dependence and QTE waveforms is currently unknown, and further investigations are therefore warranted.

**Conclusions**

Non-conventional diffusion encoding can probe microstructural features beyond those available with SDE. The preferred platform for in vivo experiments based on, for example, DDE, has been 3 T systems with high-performance gradients. In this study, we have demonstrated that QTE is technically feasible at a wide range of MRI systems, with main magnetic fields between 1.5 and 7 T, and gradient amplitudes as low as 33 mT/m. We have also demonstrated whole-brain DIVIDE in 8 minutes. The implementation was facilitated by efficient encoding waveforms and improved sampling protocols. By enabling QTE and DIVIDE at a wide range of systems, we expect that such methods may be more broadly used, which could open up new and exciting venues for dMRI research.

**Appendix – Protocol design**

The design of the protocol followed three steps. Step (i), determine the minimal and maximal b-values ($b_{min}$ and $b_{max}$). An upper bound of the maximal b-value ($b_{max}$) was imposed by the assumption that the phase dispersion is approximately Gaussian (Lasič et al., 2014). This assumption is approximately valid if the signal is not attenuated below 10%, that is, $S(b_{max})/S_0 = \exp(-b \cdot MD) > 0.1$ (Topgaard and Söderman, 2003). Assuming a worst case scenario where STE renders monoexponential signal decay, we set the upper limit by requiring that

$$b_{\max} < -\frac{\ln(0.1)}{\mathrm{MD}}. \qquad \text{A. 1}$$

Thus, we get $b_{\max} < 2.3$ ms/µm$^2$ assuming MD ≈ 1 µm$^2$/ms in brain tissue. The optimal value of $b_{\max}$ depends on many factors, but in the context of DKI, it has been suggested that a reasonable value is $b_{\max} = 2$ ms/µm$^2$ (Jensen and Helpern, 2010, Hui and Jensen, 2015), which we adopted. The minimal b-value ($b_{\min}$) is commonly set to zero, however, to reduce the influence of incoherent intra-voxel motion (Le Bihan et al., 1986), we set $b_{\min} = 0.1$ ms/µm$^2$.

Step (ii), determine the TE, TR, and the number of samples. The TE was minimized under the constraint that the necessary $b_{\max}$ was attainable by optimizing a waveform that produces the target b-tensor, and b-value, in the shortest time. This limit is typically set by the STE waveform. Several waveforms for STE exist (Mori and van Zijl, 1995, Wong et al., 1995, Moffat et al., 2005), however, we employed the optimized waveforms proposed by Sjölund et al. (2015). Subsequently, TR was determined from the TE and the number of slices.

Step (iii), distribute the samples in terms of sampling directions, b-values and tensor shapes. The total number of signal samples was NOS = $T_{\mathrm{tot}}$/TR, where $T_{\mathrm{tot}}$ is the total scan time. The samples were distributed according to two considerations. First, we assume that the accuracy of the estimated $V_{\mathrm{I}}$ and $V_{\mathrm{A}}$ are equally important, thus the total number of samples were evenly distributed across LTE and STE. Second, the samples were distributed across three equidistant b-shells. If the requirement on rotationally invariant powder signal was fulfilled in each shell, an additional shell was added until no more samples remained. This was done iteratively, and the b-shells were always equidistant.

The minimum number of diffusion encoding directions required to yield a rotation invariant powder signal was estimated from simulations, according to the preliminary report by Szczepankiewicz et al. (2016c). Briefly, the simulations investigated the effects of rotation when using LTE in a system comprising a single diffusion tensor (**D**). The powder averaged signal ($\bar{S}$) was calculated as the average diffusion-weighted signal along a number ($n_{\mathrm{dir}}$) of diffusion encoding directions (**n**, |**n**| = 1), according to

$$\bar{S}(b) = \frac{1}{n_{\mathrm{dir}}} \sum_{i=1}^{n_{\mathrm{dir}}} S_0 \exp(-b \cdot \mathbf{n}_i \mathbf{D} \mathbf{n}_i^T), \qquad \text{A. 2}$$

where $\mathbf{n}_i$ is the $i^{\mathrm{th}}$ direction in a set. This was done for 512 rotations of **D**, which yielded as many realizations of the powder averaged signal, denoted $\bar{S}_j(b)$. The rotational variance was quantified in terms of the coefficient of variation (CV) of the powder averaged signal across all rotations, according to

$$\mathrm{CV}(b) = \frac{\mathrm{SD}\left(\bar{S}_j(b)\right)}{\mathrm{E}\left(\bar{S}_j(b)\right)}, \qquad \text{A. 3}$$

where E(·) and SD(·) are the operators for the mean and standard deviation, respectively. The CV was computed for $n_{\mathrm{dir}} \in [1, 64]$. Each set of directions was independently optimized through electrostatic repulsion on the half-sphere (Jones et al., 1999, Leemans, 2009) where the optimization was repeated $10^3$ times with random initialization to avoid sub-optimal direction sets (Cook et al., 2007). The threshold for rotation invariance was set to CV < 1%. The minimum number of directions necessary to meet this condition ($n_{\min}$) was calculated as a function of FA ∈ [0, 1] and $b$·MD ∈ [0, 4]. The necessary number of directions are presented in Fig. A. 1. For example, coherent white matter where MD ≈ 1 µm$^2$/ms and FA ≈ 0.9 requires $n_{\min} = 16$ at $b = 2$ ms/µm$^2$. For low b-values $n_{\min}$ approaches 1, however, we set a minimum of at least 6 directions in each b-shell to reduce the potential influence from signal outliers. Note that a 'minimal' protocol using 3, 4 and 5 b-shells with linearly spaced $b \in [0.1, 2]$ ms/µm$^2$ require at least 29, 36 and 46 LTE samples, respectively. For reference, assuming that TR = 5 s, and with equal number of LTE and STE samples per b-shell, this amounts to $T_{\mathrm{tot}} \approx$ 5, 6, and 8 minutes.

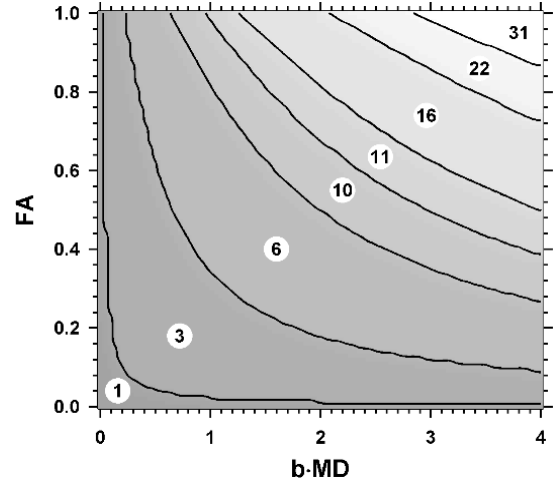

Fig. A. 1 | Minimal number of directions required ($n_{\min}$) to yield a rotationally invariant powder averaged signal. The $n_{\min}$ can be read from the figure for combinations of the tissue fractional anisotropy (FA) and attenuation factor ($b$·MD), where the numbers in the circles show $n_{\min}$. As expected, higher anisotropy and attenuation both demand a larger number of diffusion encoding directions. Note that rotational invariance is here assumed to be CV < 1%.